\documentclass[fleqn,usenatbib]{mnras}

\usepackage{mathptmx}

\usepackage[T1]{fontenc}
\usepackage{ae,aecompl}

\usepackage{graphicx}	
\usepackage{amsmath}	
\usepackage{amssymb}	
\usepackage{afterpage}
\usepackage{multirow}
\usepackage{subfigure}
\usepackage{csvsimple}	
\usepackage{pdflscape}	
\usepackage{xcolor}

\newcommand{\hersc}{{\it Herschel}}
\newcommand{\spitz}{{\it Spitzer}}
\newcommand{\chan}{{\it Chandra}}
\newcommand{\ppmap}{{\sc ppmap}}
\newcommand{\xmm}{{\it XMM-Newton}}
\newcommand{\suz}{{\it Suzaku}}
\newcommand{\HII}{H{\sc ii}}

\newcommand\ppmapRGBtempA{20}
\newcommand\ppmapRGBtempB{30}
\newcommand\ppmapRGBtempC{40}
\newcommand\ppmapRGBtempD{61}

\newcommand\ppmapDustMass{16.7}

\newcommand\jameskappa{0.56}

\graphicspath{{Figures/}}

\title[Dust Devil]{A Galactic Dust Devil: far-infrared observations of the Tornado Supernova Remnant candidate}

\author[H. Chawner et al.]{H. Chawner$^{1}$\thanks{E-mail: ChawnerHS@cardiff.ac.uk},
A.D.P. Howard$^{1}$,
H.L. Gomez$^{1}$,
M. Matsuura$^{1}$,
F. Priestley$^{1}$,
\newauthor
M. J. Barlow$^{2}$,
I. De Looze$^{2, 3}$,
A. Papageorgiou$^{1}$,
K. Marsh$^{4}$,
M.W.L. Smith$^{1}$,
\newauthor{
A. Noriega-Crespo$^{5}$,
J. Rho$^{6}$, and
L. Dunne$^{1}$}
\\
$^{1}$School of Physics and Astronomy, Cardiff University, Queens Buildings, The Parade, Cardiff, CF24 3AA, UK\\
$^{2}$Department of Physics and Astronomy, University College London, Gower Street, London WC1E 6BT, UK\\
$^{3}$Sterrenkundig Observatorium, Ghent University, Krijgslaan 281 S9, B-9000 Gent, Belgium\\
$^{4}$IPAC, Caltech, 1200 E California Blvd, Pasadena, CA 91125, USA\\
$^{5}$Space Telescope Science Institute, 3700 San Martin Drive, Baltimore, MD 21218, USA\\
$^{6}$SETI Institute, 189 N. Bernardo Ave, Suite 100, Mountain View, CA 94043, USA\\
}

\date{Accepted XXX. Received YYY; in original form ZZZ}

\pubyear{2017}

\begin{document}
\label{firstpage}
\pagerange{\pageref{firstpage}--\pageref{lastpage}}
\maketitle

\begin{abstract}
	We present complicated dust structures within multiple regions of the candidate supernova remnant (SNR) the `Tornado' (G357.7$-$0.1) using observations with \spitz\ and \hersc.
	We use Point Process Mapping, \ppmap, to investigate the distribution of dust in the Tornado at a resolution of $8^{\prime \prime}$, compared to the native telescope beams of $5-36^{\prime \prime}$. We find complex dust structures at multiple temperatures within both the head and the tail of the Tornado, ranging from 15 to 60\,K. Cool dust in the head forms a shell, with some overlap with the radio emission, which envelopes warm dust at the X-ray peak.
Akin to the terrestrial sandy whirlwinds known as `Dust Devils', we find a large mass of dust contained within the Tornado.
We derive a total dust mass for the Tornado head of \ppmapDustMass\,$\rm M_{\odot}$, assuming a dust absorption coefficient of $\kappa_{\rm 300} =$\jameskappa$\,\rm m^2\,kg^{-1}$, which can be explained by interstellar material swept up by a SNR expanding in a dense region.
	The X-ray, infra-red, and radio emission from the Tornado head indicate that this is a SNR. The origin of the tail is more unclear, although we propose that there is an X-ray binary embedded in the SNR, the outflow from which drives into the SNR shell. This interaction forms the helical tail structure in a similar manner to that of the SNR W50 and microquasar SS433.

\end{abstract}


\section{Introduction}
\label{sec:intro}

\begin{figure}
	\includegraphics[width=\linewidth]{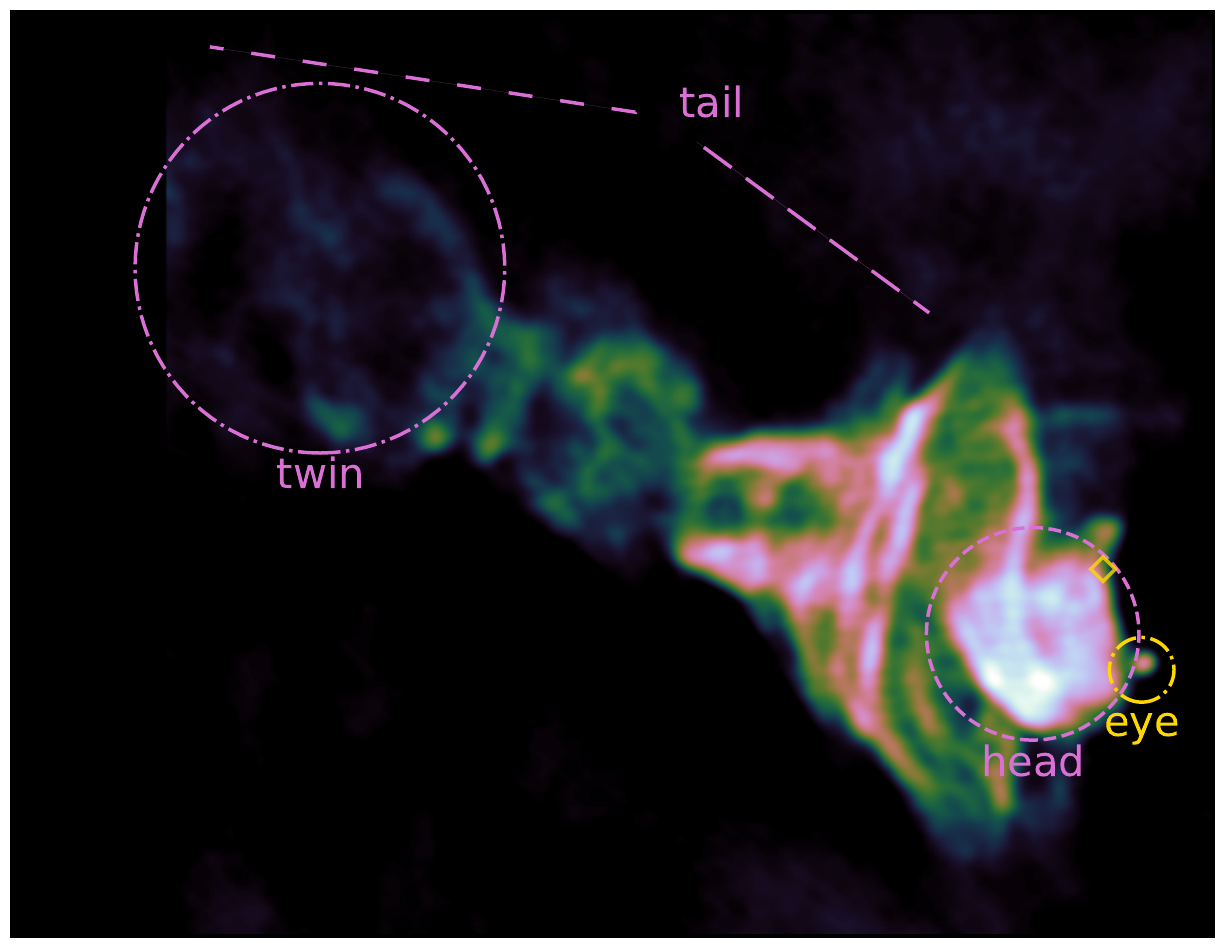}
	\caption{1.4\,GHz VLA continuum image of the Tornado \citep{Brogan2003}. The tail, head, and eye are indicated, as well as the X-ray `twin' of the head, detected by \citet{Sawada2011}. Like \citet{Gaensler2003}, we define the head as the region from which both X-ray and radio emission are strongly detected. The gold diamond indicates the location of an OH (1720\,MHz) maser.}
	\label{fig:RoadMap}
\end{figure}

\begin{figure*}
	\includegraphics[width=0.95\linewidth, trim=1cm 2.4cm 0cm 1.5cm]{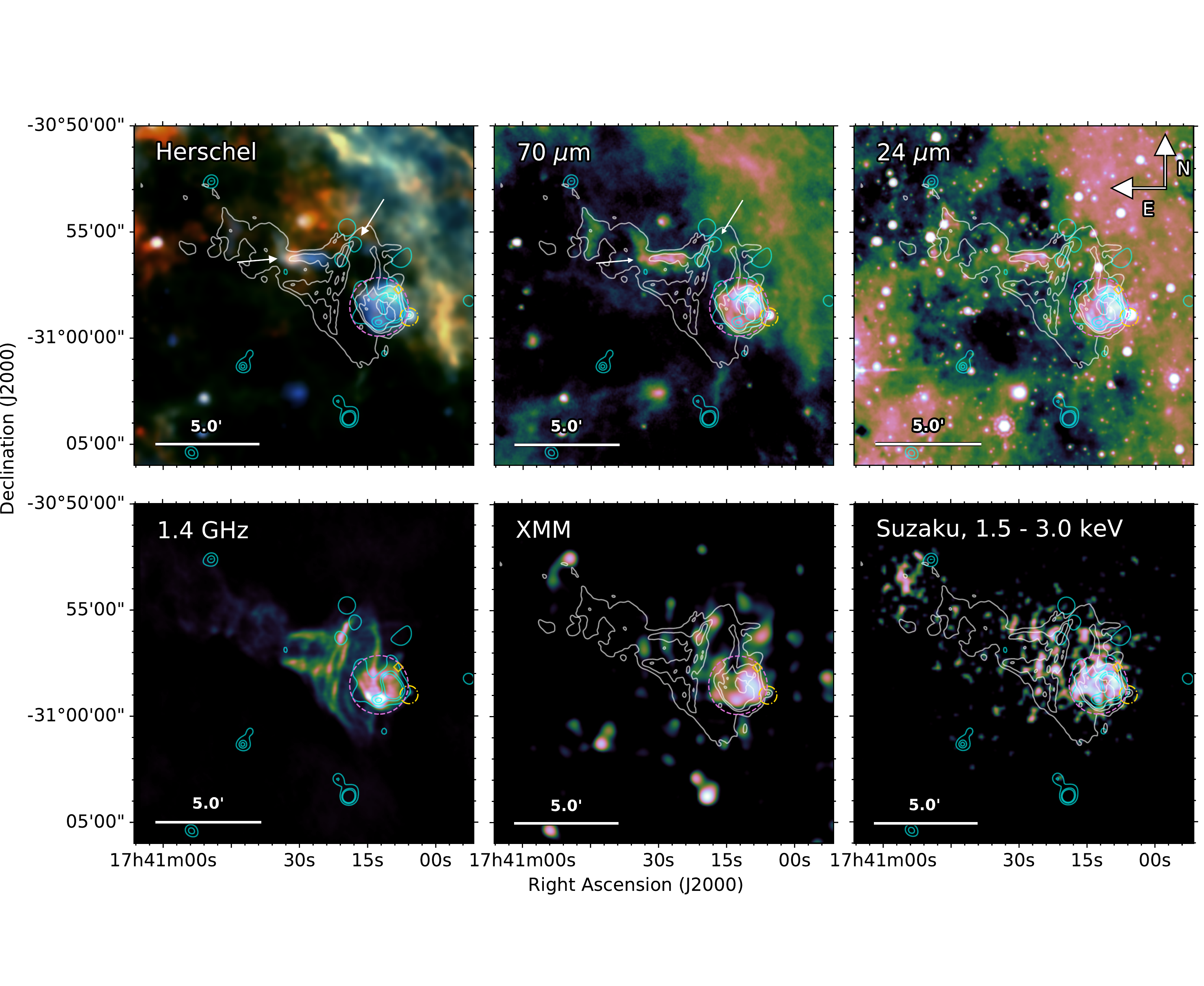}{}
	\caption{G357.7$-$0.1, The Tornado at FIR, radio, and X-ray -
	{\it top left:} \hersc\ three colour image made by combining the 70 (blue), 160 (green) and 250\,$\mu$m (red) images,
	{\it top middle:} \hersc\ 70\,\micron\ image,
	{\it top right:} \spitz\ MIPS 24\,\micron\ image,
	{\it bottom left:} 1.4\,GHz VLA image,
	{\it bottom middle:} \xmm\ X-ray image smoothed to 0.5$^{\prime \prime}$ pixels (kindly provided by B. Gaensler et al. private communication), and
	{\it bottom right:} \suz\ 1.5\,--\,3.0\,keV X-ray smoothed continuum image.
	The white contours show the radio emission (1.4\,GHz VLA) and the cyan contours show X-ray emission (\xmm).
	We detect dust emission across all \hersc\ wavebands from the `head' of the Tornado, within the pink circle. We also detect FIR emission from the `tail' of the Tornado, and from a fainter filament extending around the head, as indicated by the arrows.
	The gold diamond indicates the location of an OH (1720\,MHz) maser.
	(For the single wavelength panels we use the {\sc cubehelix} colour scheme, \citet{Green2011}.)
	}
	\label{fig:G357.7-0.1Image}
\end{figure*}

Known as `the Tornado', G357.7$-$0.1 (MSH 17-39) is an unusual SNR candidate at a distance of 11.8\,kpc \citep{Frail1996}, comprising a `head', `tail', and `eye'  (Fig.~\ref{fig:RoadMap}).
The head appears as a shell- or ring-like feature in the radio \citep{Shaver1985}, and a `smudge' or diffuse clump with a southern peak in the X-ray, with \suz\ \citep{Sawada2011} and \chan\ \citep[see Fig.~\ref{fig:G357.7-0.1Image} of][]{Gaensler2003} respectively. A larger extended radio shell/filamentary structure exists around the head, with an elongated tail. Finally, a compact and bright radio source seen to the west of the head at $\alpha = 17^\text{h}40^\text{m}05.9^\text{s}, \delta = -30^\circ59^\prime00''$ (J2000) is the so-called eye of the Tornado, which is an isolated core embedded in a foreground H{\sc ii} region \citep{Brogan2003, Burton2004}, unrelated to the SNR structure.

Its highly unusual structure has led to various origin theories for the Tornado.
From early days, the head of the Tornado has been attributed to a SNR with its radio power law index following synchrotron emission, its non-thermal radio emission, and its strong polarisation \citep[e.g.][]{Milne1979, Shaver1985, Becker1985}, and later its X-ray emission power-law index \citep{Yusef-Zadeh2003, Gaensler2003}. These properties led \citet{Gaensler2003} to propose that the Tornado is a shell or mixed morphology SNR, as described by \citet{Rho1998}.
The radio head of the Tornado (which is brightest in the south-west part of the `shell' with a peak in the north-west) can be attributed to limb brightened emission due to the interaction with a molecular cloud \citep{Gaensler2003}. Indeed, shocked $\rm H_2$ gas detected along the north-western edge of the head \citep{Lazendic2004}, and the presence of multiple OH masers \citep{Frail1996,Hewitt2008} both support this scenario.
Unshocked CO emission is found from a cloud to the north-west slightly offset from shocked $\rm H_2$, suggesting that there is a dense molecular cloud ($\rm n_H \sim 10^4-10^6\,\rm cm^{-3}$) which could decelerate the shock wave on this side \citep{Lazendic2004}.
However, it is difficult to explain the shape of the large filamentary structures in the tail (Fig.~\ref{fig:G357.7-0.1Image}) with a mixed morphology SNR.
In this scenario, the X-ray emission from the head (detected with \chan) originates from the SNR interior, i.e. interior to the limb brightened radio shell \citep{Gaensler2003}. Outside the head region, \citet{Shaver1985} suggests that the partial helical/cylindrical radio filaments could be the result of an equatorial supernova outburst, or the SN exploded at the edge of dense circumstellar shell \citep{Gaensler2003}, or a pre-existing spiral magnetic field structure \citep{Stewart1994}.

Another explanation is that the helical tail is a structure originating from jets of an X-ray binary, as seen in the SNR W50 \citep{Shaver1985, Helfand1985a, Stewart1994}. In that system, over the course of 20\,kyr and several episodes of activity, precessing relativistic jets of the X-ray binary SS443 have shaped the SNR within which it is found \citep[e.g.][]{Begelman1980, Goodall2011}. This has resulted in a huge nebula (208\,pc across) which has a circular radio shell (with a 45\,pc radius) from the expanding SNR, and lobes extending to 121.5 and 86.5\,pc to the east and west respectively formed by outflows.
Radio observations of the Tornado show some symmetry, with flared ends and a narrower central region \citep{Caswell1989}, and \citet{Sawada2011} suggested the presence of an X-ray `twin' to the head at the far end. This has lead to the theory that the Tornado is an X-ray binary, with a powering source near to the centre of the radio structure, and bipolar jets which interact with ISM at either end, forming the head and its `twin'.

However, a compact object powering the Tornado system has not yet been detected \citep{Gaensler2003},
although \citet{Sawada2011} argued that a central powering source with an active past may now be in a quiescent state and is too faint to detect in X-ray emission.
Another proposed idea is that the Tornado is a pulsar wind nebula powered by a high-velocity pulsar \citep{Shull1989}; however, the spectral slope required to explain the X-ray emission is too steep \citep{Gaensler2003}.
Currently, the origin of the highly unusual shaping observed in the Tornado is still under debate.

SNRs are considered to play an important role in the dust processes in the ISM, by creating freshly formed ejecta dust and destroying pre-existing interstellar dust. Indeed, dust thermal emission is widely detected in SNRs in the mid- and far-infrared (MIR and FIR) regime \citep{Dunne2003,William2006,Rho2008,Barlow2010,Matsuura2011,Temim2012,Gomez2012b,DeLooze2017,Temim2017,Rho2018,Chawner2019,DeLooze2019}. As SNRs plough through surrounding interstellar dust clouds, they form a shell-like structure, whereas ejected material is found in a compact emission source in the center of the system \citep{Barlow2010,Indebetouw2014}. Using MIR to FIR images of the region from the {\it Spitzer Space Telescope} \citep[\spitz,][]{Werner2004} and the {\it Herschel Space Observatory} \citep[\hersc,][]{Pilbratt2010}, \citet{Chawner2020} reported the discovery of thermal emission from dust in the head and tail of the Tornado (see Section~\ref{sec:FIRSurvey}). This paper examines the unusual morphology of dust emission in the SNR candidate, the Tornado.

\section{The Infrared View of the Tornado} \label{sec:FIRSurvey}

\subsection{Observations}
The \hersc\ data used to discover dust emission in the Tornado is from the HiGal survey \citep{Molinari2011,Molinari2016}, which covered $360^{\circ}$ in longitude and $\mid b \mid \leq 1$ and includes data from 70\,--\,500\,$\mu$m.  Data processing is described in detail in \citet{Molinari2016} and pipeline-reduced and calibration corrected fits files are available to the community via the native HIPE reduction pipeline.
Zero-point calibrations for the \hersc\ SPIRE observations were already applied prior to data acquisition. The \hersc\ PACS zero-point offsets were corrected by comparing the observations to synthetic observations produced from the {\it Planck} foreground maps \citep{Planck2016}, and the 100\,\micron\ {\em IRAS}~ IRIS data\footnote{The zero-point corrections adopted for the G357.7-0.1 region are: 66.1\,MJy$/$sr and 454.1\,MJy$/$sr for 70\,\micron\ and 160\,\micron\ respectively.}. This method is similar to that described in e.g. \citet{Bernard2010, Lombardi2014, Abreu-Vicente2016}.
\spitz\ 24\,\micron\ data was available via the IRSA archive. The MIR-submm images of the Tornado are presented in Fig.~\ref{fig:G357.7-0.1Image} (and Fig.\,\ref{fig:HeadMultiband}), where the well known features are marked by a magenta circle (the head), arrows (the tail) and a gold circle (the H{\sc ii} region, the eye). The tail is brightest in two prong-like structures east of the head.

Fig.~\ref{fig:G357.7-0.1Image} also compares the IR images with other physical tracers. We make use of the 1.4\,GHz VLA radio image (with spatial resolution of $14^{\prime \prime} \times 11^{\prime \prime}$, \citealt{Brogan2003})
and X-ray data from the EPIC camera on board \xmm\ (kindly provided by B. Gaensler et al. private communication), with an energy range 0.15-15\,keV and spatial resolution of $6^{\prime \prime}$. As the source was only weakly detected in the EPIC MOS detector, here we present data from the PN detector only.
We use \xmm\ rather than \chan\ as we are only interested in the comparison of structures rather than absolute flux or spectral variations. Furthermore, the diffuse source concentrated at the south of the head previously detected with \chan\ \citep{Gaensler2003} is very faint and requires significant smoothing to bring out the signal; \xmm\ may ultimately be more sensitive to diffuse emission given its coarse angular resolution compared to \chan. X-ray observations from \suz\ \citep{Sawada2011} suggest faint diffuse X-ray emission across the head of the Tornado, in close agreement to the structures observed in the \xmm\ image (Fig.~\ref{fig:G357.7-0.1Image}). We note that the distribution of X-rays as seen in the \xmm\ image suggest a shell-like X-ray structure with some emission in the south which may lie interior to the shell (i.e. potentially originating from ejecta; see also the peak in the smoothed \chan\ image of \citealp{Gaensler2003}).

\subsection{Comparison of Tracers} \label{sec:TracerComparison}
Although this region is confused by dust in the interstellar medium (ISM) in the FIR, we detect clear emission from dust at the location of the head and tail of the SNR in all \hersc\ wavebands, as shown in Fig.~\ref{fig:G357.7-0.1Image} and Fig.\,\ref{fig:HeadMultiband}, though the poorer resolution at 350 and 500\,\micron\ makes it more difficult to distinguish the emission from unrelated structure along the line of sight.
At 70 and 160\,\micron, the shell-like structure is clearly seen in the head, and correlates spatially with the radio and overlaps with X-ray. This is also confirmed in the {\it Spitzer} 24\,$\mu$m image (Fig.~\ref{fig:G357.7-0.1Image}). The brightest peak in the MIR and FIR (to the north and north-west) is opposite to that seen in the radio emission, and is located towards the OH (1720\,MHz) maser, where shock-heated $\rm H_2$ is also bright \citep{Lazendic2004}. This dust feature appears confined within the radio contours, and is significantly brighter than the ambient dust seen further north-west where the interacting IS cloud is located (as traced by molecular CO emission; \citealp{Lazendic2004}) so there is no doubt that this is associated with the emission structures responsible for the radio and X-ray (i.e. shocked gas). The fainter southern peak in the X-ray emission correlates with two radio peaks, and the bright X-ray feature to the west coincides with the brightest 24\,$\mu$m emission and fainter radio.

Outside of the head, we detect warm dust in the unrelated H{\sc ii} region. We also detect faint 70\,\micron\ emission that appears to correspond to one of the large radio filaments extending around the eastern side of the head. Dust emission from the tail is also seen at 24\,--\,160\,$\mu$m. 
Similar to the head, we see evidence of an anti-correlation between the radio and FIR in the tail, the FIR correlates with the upper, fainter of the two radio prongs, indicated by an arrow in Fig.~\ref{fig:G357.7-0.1Image}. At longer \hersc\ wavelengths, we see a bright structure at the eastern end of the tail which may be associated with the Tornado, although this is difficult to distinguish from interstellar dust due to the level of confusion in this region. We do not discuss this source further.

\section{Investigating the dust structures in the Tornado}
\label{sec:methods}

In the previous Section, we discussed the presence of dust in the SNR G357.7-0.1, `the Tornado' (Fig.~\ref{fig:G357.7-0.1Image}). Here we investigate the dust properties in this source further using the point process mapping technique, \ppmap.  This technique produces maps of differential dust column density for a grid of temperatures \citep{Marsh2015,Marsh2017}.
Observations are taken at their native resolution, avoiding data loss through degrading to a common angular scale, and are deconvolved with circularly average instrument beam profiles, using the point spread function information, to achieve maps of dust mass at a high resolution.
Finely sampled colour corrections, derived from the \spitz\ MIPS and \hersc\ PACS and SPIRE response functions, are applied to the model fluxes, as a function of temperature and wavelength.

The \ppmap\ procedure is described in full in \citet{Marsh2015,Marsh2017} and its application to investigating the dust properties in pulsar wind nebulae can be found in \citet{Chawner2019}. In brief, \ppmap\ uses an iterative procedure based on Bayes' theorem to estimate a density distribution of mass in the state space $(x,~ y,~ T,~ \beta)$ where $x$ and $y$ are spatial co-ordinates, $T$ is the dust temperature and $\beta$ is the dust emissivity index (the power law slope that characterises how the dust opacity varies with wavelength).
Throughout the procedure, \ppmap\ acts in the direction of minimising the reduced $\chi^2$, derived from the sums of squares of deviations between the observed and model pixel values over each local region, after dividing by the number of degrees of freedom. These are estimated by comparing the estimated properties of each tile with a modified black-body model of the form:

\begin{equation} \label{eqn:Greybody}
	\centering
	F_\lambda = \frac{M_{\rm dust} B_\lambda (T) \kappa_\lambda}{D^2},
\end{equation}

where $F_\lambda$ is the flux at a given wavelength, $M_{\rm dust}$ is the mass of dust, $B_\nu (T)$ is the Planck function at temperature $T$, $\kappa_\lambda$ is the dust mass absorption coefficient, and $D$ is the distance to the source, which is $\sim$12\,kpc in this case. The variation of $\kappa_\lambda$ at different wavelengths depends on the value of $\beta$ as $\kappa_\lambda = \kappa_{\lambda_0}(\lambda/\lambda_0)^{-\beta}$. 
We adopt $\kappa_{\rm 300} =\jameskappa \rm m^2\,kg^{-1}$ \citep{James2002} in the \ppmap\ analysis.

The process is applied to a multi-band map field to estimate the column density over a range of temperatures. \ppmap\ provides additional information over the standard modified blackbody technique used to derive dust masses because it (i) does not assume a single dust temperature along the line of sight through each pixel, (ii) uses point spread function information to create column density maps without needing to smooth data to a common resolution, and (iii) although it first makes the assumption that the dust is optically thin, it can check this retrospectively. \ppmap\ requires an estimate of the noise levels for each band which describes the pixel-to-pixel variation.  Here, this was derived from background subtracted \spitz\ and \hersc\ images using the standard deviation of pixels within apertures placed in quiet regions (minimal variation in foreground emission) near the source. This gives noise estimates of 2.18, 5.47, 11.87, 4.10, 1.72, and 0.48\,MJy\,sr$^{-1}$ for the 24, 70, 160, 250, 350, and 500\,\micron\ bands respectively, which are assumed to be uniform across the entire map.

\subsection{Applying \ppmap\ to the Tornado}

We initially selected 12 temperature bins centred at temperatures equally spaced in ${\rm log}(T)$ ranging from 20 to 90\,K (guided by our previous analysis of SNRs in C19), we assumed a fixed value for the dust emissivity index, $\beta =2$, which is typical for silicate ISM dust \citep{Planck2016beta}. If we were to assume a carbonaceous dust with $\beta$ of 1.0 to 1.5 the estimated dust temperatures would likely be higher. As we did not find any related dust at the location of the head in any temperature bins >\,70\,K, we re-ran the grid for temperatures ranging from 15 to 70\,K.

In our first runs of \ppmap, we found that the iterative procedure did not converge to sensible fits (verified by checking the \ppmap\ $\chi^2$ statistic in each band), even with hundreds of thousands of iterations. This was due to \ppmap\ attempting, and failing, to converge to a solution for the bright point sources, presumably stars with temperatures much higher than 90~K, in the 24\,$\mu$m image (and to a lesser extent in the 70\,$\mu$m image).   To resolve this, we masked the bright point sources near the Tornado (replacing their pixels with a local average level in the image) and we artificially increased the noise for the 24\,$\mu$m map by a factor of 10; this effectively stops \ppmap\ from trying to over-fit the 24\,$\mu$m band and down-weights the importance of the 24\,$\mu$m in the iterative procedure.  This may act to slightly reduce any dust temperatures fit by \ppmap, though in practice we found that it did not affect our results.

The Tornado is in a highly confused region due to its location close to the Galactic centre (Fig.~\ref{fig:G357.7-0.1Image}). To determine the effect of any potential contamination from unrelated dust along the line of sight, we ran our \ppmap\ grid (the original 20\,--\,90\,K run) on the Tornado without any background subtraction, and then again, after accounting for background emission. In the former scenario, the results indicate that dust structures exist in the head of the Tornado at temperatures of 20-23\,K with a warmer dust component in the north-western part of the head at 26\,K, where the source is believed to be interacting with a molecular cloud \citep{Frail1996,Lazendic2004,Hewitt2008}.  These cold dust temperatures are very similar to general interstellar dust, and the narrow range of temperatures suggest this region is contaminated by unrelated background emission.

For \ppmap\ to converge in a reasonable time we must subtract the background from the maps.
First we mask bright, unrelated sources as above, as well as the Tornado head and tail, and several high signal-to-noise regions to avoid overestimating the background. The images are then convolved with a 100\,$^{\prime\prime}$ FWHM Gaussian profile, providing background maps smoothed to a scale comparable to the Tornado head. The background maps are subtracted from the original zero-point calibrated maps (with the two bright sources masked). Running \ppmap\ with the resulting maps gives reduced-$\chi^2$ values of 0.3, 2.0, 11.0, 9.0, 4.0, and 128.0 for 24, 70, 160, 250, 350, and 500~\micron \footnote{These are average reduced-$\chi^2$ estimated for the entire map at the end of the \ppmap\ run. As such they can be greatly influenced by variations in noise across the map, as well as regions which are not fit well, including edges (which are sampled less frequently throughout the \ppmap\ procedure) and areas which may be optically thick or have a temperature outside of the given range. }
We find that the overall level of the background-subtracted images is negative, implying the method of background subtraction used is too aggressive.  To account for this, we took the background-subtracted maps, estimated the mean negative offset for the whole region at each waveband (again masking the Tornado) and added this back on to the image in an attempt to bring the maps back to a zero level.   Hereafter we call this the zero mean background-subtracted method. Running these images through \ppmap\ the resulting dust temperatures and components are markedly different to the non-background-subtracted case: dust structures are observed at a wider range of temperatures (from 20 - 60\,K) with the north-western dust feature peaking at 30\,K. The background subtraction has resulted in the dust components in the head being attributed to warmer dust, as expected.  Note that these warmer dust components agree with the dust structures that peak in the original {\it Herschel} maps peaking at 70\,$\mu$m.
The resulting \ppmap\ reduced $\chi^2$ values are 0.6, 2.2, 6.9, 11.7, 22.5 and 37.6 suggesting the overall fit is formally better than the previous case. 
The high $\chi^2$ values for the longer wavebands are most likely due to underestimating the $\sigma$ value, because small scale ISM variations cannot be captured by a large beam, although increasing the noise level constrains \ppmap\ less, giving more unreliable results across all bands.

The above tests suggest that \ppmap\ is sensitive to whether the background diffuse interstellar level is subtracted from the maps or not, particularly important in this case due to the high level of confusion in this region.  To try and qualitatively discriminate between the tests, we created synthetic MIR-FIR observations based on the \ppmap\ outputs for the three scenarios above, and compared them to the original {\it Spitzer} and {\it Herschel} images.   In each case, the original dust emission features seen in the head of the Tornado were recovered well in the synthetic \ppmap\ MIR-FIR images.  The zero mean background-subtracted method provided the closest match to the original features (see Appendix~\ref{app:ppmaptests}), recovering the complex dust emission structures observed within the head (see the following Section for more information). We therefore use the \ppmap\ results based on this method from now on.

Finally we note that synchrotron emission in SNRs can be a significant contributor to the FIR flux \citep{Dunne2003, DeLooze2017, Chawner2019}. As this typically varies as a power law with flux $S_{\nu} \propto \nu^{-\alpha}$ where $\alpha$ is the spectral index, we can directly estimate the contribution of synchrotron emission to our FIR bands.
Prior to running \ppmap\ we subtract the synchrotron contribution which is estimated by extrapolating from the flux we measure from the 1.4\,GHz VLA image \citep{Becker1985a,Green2004}, assuming $\alpha=\,-0.63$ for the head \citep{Law2008}.
We find that the synchrotron contribution to the SNR head is in the range of only 0.03\,--\,2.06\,per\,cent of the total flux for our MIR--FIR wavebands in the head, as measured on the original \hersc\ maps\footnote{we note that this calculation may underestimate the synchrotron contribution to the IR fluxes since our integrated flux for the total SNR (head and tail) derived from the 1.4\,GHz radio image using an aperture $\alpha\,=\,17^\text{h}40^\text{m}29^\text{s}, \delta\,=\,-30^\circ58^\prime00''$ with a 8$^{\prime}$ radius, gives 80 and 70\,per\,cent of the flux derived from the single dish measurements of \citet{Green2004} and \citet{Law2008} respectively (scaled to the same frequency). This may, in part, explain the larger $\chi^2$ value at 500\,\micron. However taking the single dish measurements would produce a maximum synchrotron contribution of 3\,per\,cent. Indeed the biggest source of contamination in the MIR-FIR aperture measurements is the background level.},
where both are measured within an aperture centred at $\alpha\,=\,17^\text{h}40^\text{m}12.4^\text{s}, \delta\,=\,-30^\circ58^\prime31.1''$ with a 79$^{\prime\prime}$ radius. We can therefore be confident that we are observing the thermal emission from dust with negligible contribution from synchrotron emission in the head.

However, the spectral index does flatten in the tail region with spectral slope varying from $-0.50\,<\,\alpha\,<\,-0.33$ \citep{Law2008} indicating that the tail electrons are more energetic than in the head. We therefore caution that there could be a higher contribution of synchrotron emission in the tail.

\subsection{Results} \label{subsec:ppmapResults}
\begin{figure}
	\includegraphics[width=\linewidth, trim = 0.3cm 0.2cm 0.3cm 0.2cm, clip]{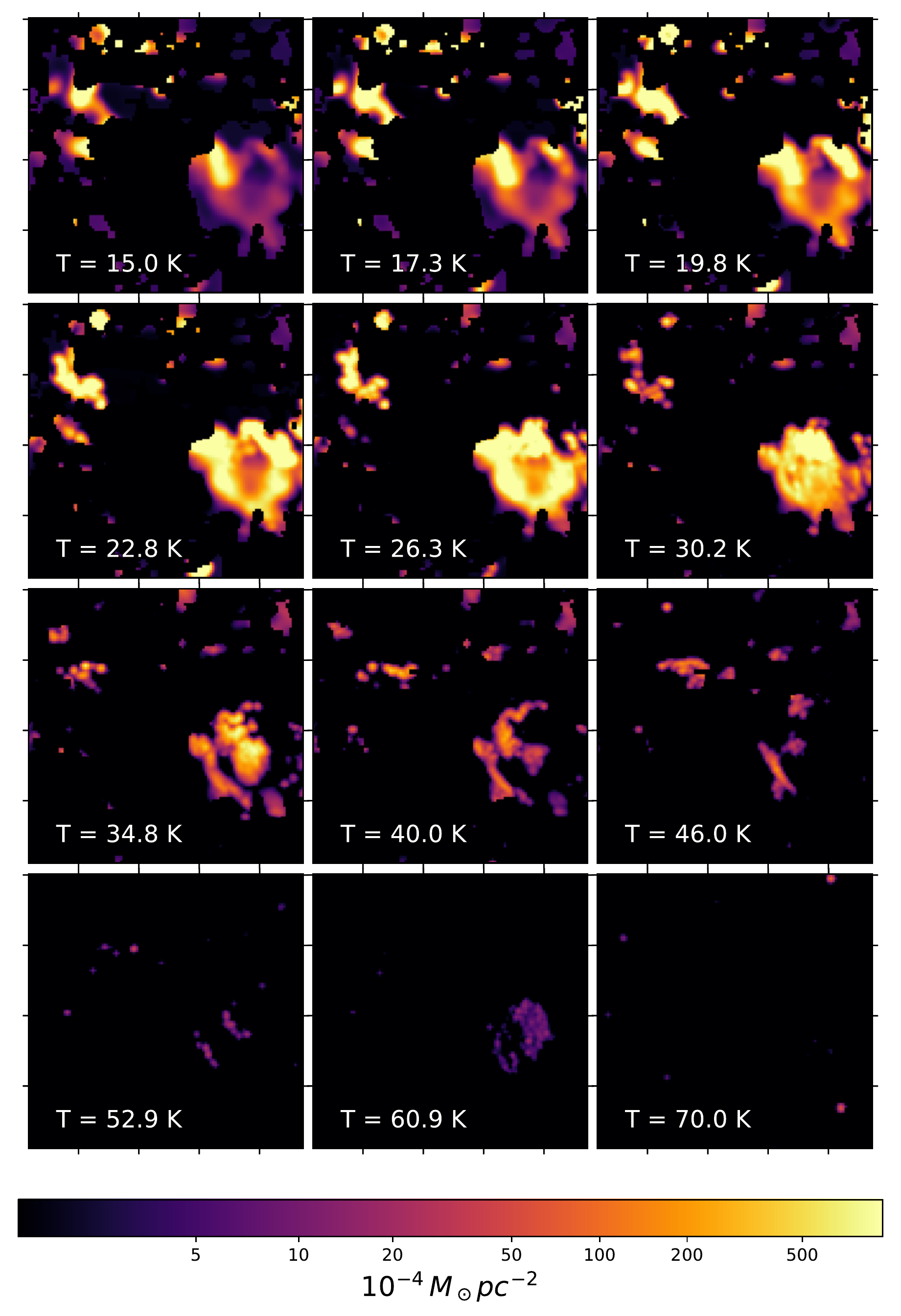}
	\caption{\ppmap\ generated maps of differential dust mass split in different temperature ranges for the Tornado. The corresponding dust temperature is indicated in the bottom-left of each panel. }
	\label{fig:tornado_temps_ppmap}
\end{figure}

\begin{figure}
	\includegraphics[width=\linewidth]{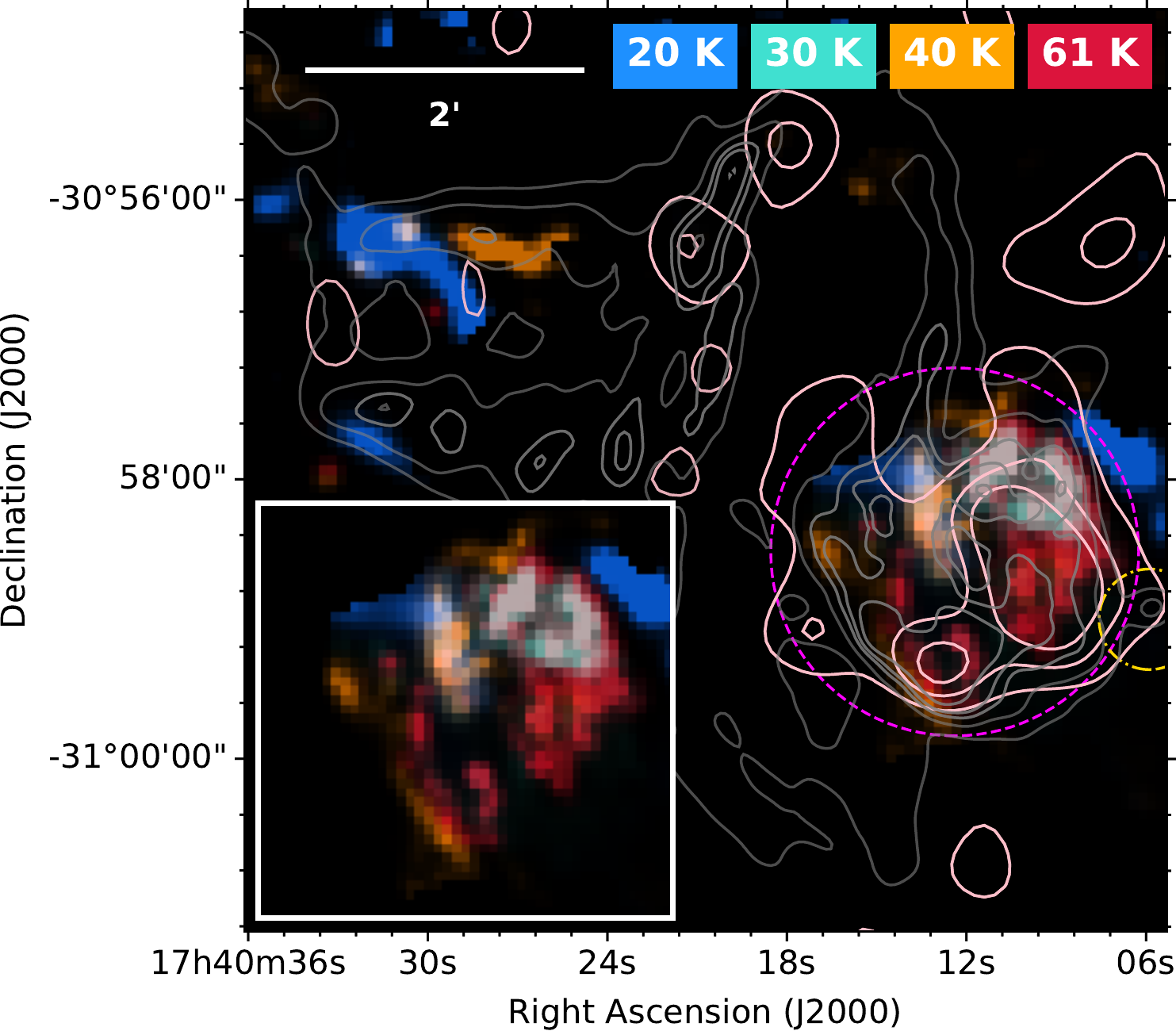}
	\caption{
	\ppmap-generated four colour map of dust mass in the Tornado created using dust temperature slices from Fig.~\ref{fig:tornado_temps_ppmap}. Colours show dust at \ppmapRGBtempA\,K (blue), \ppmapRGBtempB\,K (cyan), \ppmapRGBtempC\,K (gold), and \ppmapRGBtempD\,K (red). Overlaid contours are from the VLA 1.4\,GHz (grey) and \xmm\ (pink) images.
	The magenta dashed circle indicates the location of the head of the remnant, and is also the aperture used to derive the dust mass. The gold dash-dotted circle is the location of the eye of the Tornado (unrelated H{\sc ii} region).
	}
	\label{fig:tornado_temps_3col_ppmap}
\end{figure}

\begin{figure}
	\includegraphics[width=1\linewidth]{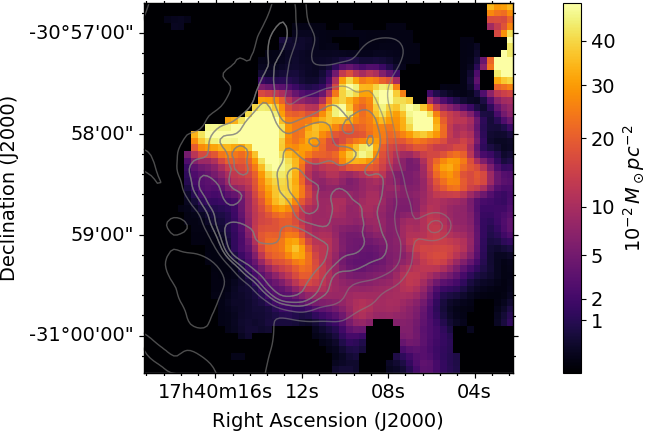}
	\caption{
	The dust mass within the Tornado head integrated across all temperature slices of Fig.~\ref{fig:tornado_temps_ppmap}, with VLA 1.4\,GHz contours (grey) overlaid.
	}
	\label{fig:tornado_total_mass}
\end{figure}

\begin{figure}
	\includegraphics[width=0.95\linewidth]{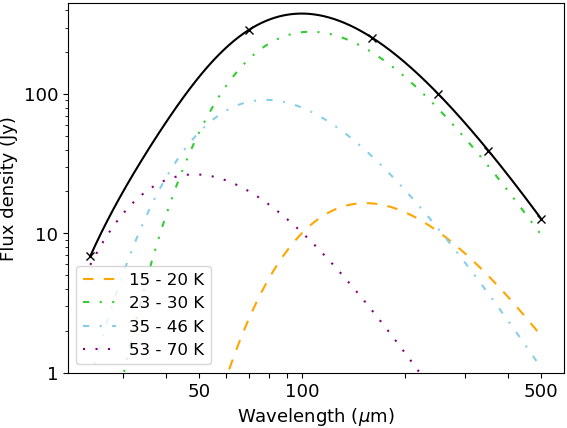}
	\caption{
	The total thermal MIR-FIR SED estimated from the \ppmap\ results of the head of the Tornado, within the magenta circle in Fig.~\ref{fig:tornado_temps_3col_ppmap}, indicating how the different temperature components shown in Fig.~\ref{fig:tornado_temps_ppmap} contribute to the thermal emission observed in the source.
	}
	\label{fig:tornado_sed}
\end{figure}

The grid of dust mass in each temperature bin for the Tornado is shown in Fig.~\ref{fig:tornado_temps_ppmap} assuming a distance of 12\,kpc \citep{Brogan2003}. Fig.~\ref{fig:tornado_temps_3col_ppmap} shows a four colour FIR image created by combining the masses in the temperature slices at \ppmapRGBtempA, \ppmapRGBtempB, \ppmapRGBtempC, and \ppmapRGBtempD\,K, and Fig.~\ref{fig:tornado_total_mass} shows the total dust mass distribution across the Tornado head. They reveal dust features observed in the \hersc\ images, but at a resolution of $\sim 8^{\prime\prime}$ compared to the native telescope beams of $5-36^{\prime\prime}$.

A temperature gradient is evident in both the head and tail.
Cool, dense dust is found towards the north-eastern head at the location of a radio filament which extends from the head towards the northern extent of the object. The filaments outside of the head were lost in background subtraction, but this suggests that they could also contain cool, dense dust.
Slightly warmer material (23\,--\,30\,K) forms a bubble around the edge of the head and around the larger X-ray peak. In Fig.~\ref{fig:tornado_total_mass} we find that the majority of the dust mass follows this bubble shape, with a relative lack of material in the central region.
Warm material (35\,--\,40\,K) fills the central region, coincident with both the large X-ray peak and the warmest dust that we observe (53\,--\,61\,K). It seems that the hot gas which emits the X-ray emission is heating the central region of the head, where we see warm, low density material.
We find a large mass of 26\,--\,30\,K dust towards the north west where interactions with a molecular cloud may be heating the dust, as well as at the same location as the smaller region of bright X-ray emission in the south east.
A filament of 35\,--\,46\,K material sits along the eastern edge of the head, with a warm 53\,K peak towards the middle, filling the radio contours at this location, as seen in Fig.\ref{fig:tornado_temps_3col_ppmap}.
In the tail we find a large mass of cool, 15\,--\,20\,K dust to the east, as well as slightly warmer, 23\,--\,30\,K material which extends further north. The temperature increases towards the west, as 35\,--\,40\,K dust fill the eastern and central contours with dense regions at the radio peaks, and 46\,K material is found further west. There is some evidence of warm dust (40\,--\,46\,K) at the X-ray and radio peak to the east of the tail, although much of this area is lost to background subtraction as it is a similar level to the surrounding ISM.

The spectral energy distribution of the head of the Tornado is shown in Fig.~\ref{fig:tornado_sed}, broken down into the different temperature components revealed by \ppmap. We derive the total dust mass in the head of the Tornado by summing the mass within the magenta circle shown in Fig.~\ref{fig:tornado_temps_3col_ppmap} across the temperature grids.
This gives a total dust mass for the Tornado head of \ppmapDustMass\,M$_\odot$ for a dust mass absorption coefficient at 300~\micron\ of $\kappa_{\rm 300} =$\jameskappa$\,\rm m^2\,kg^{-1}$ \citep{James2002}. If we only sum the contribution from dust structures with T$_d$>17\,K we obtain a dust mass of 14.8\,M$_\odot$, and 4.0\,M$_\odot$ of mass originates from dust hotter than $30$K.

\section{Dust Grain Properties} \label{sec:dustgrains}
\begin{figure}
	\centering
	\includegraphics[width=\linewidth]{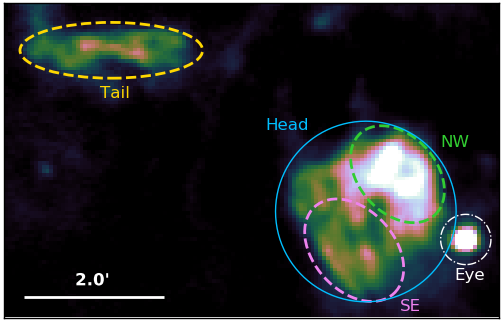}
	\caption{Tornado head and tail region at 70\,\micron. The shapes indicate regions from which we detect FIR emission and within which we compare the flux ratios in Figs.\,\ref{fig:col_plots1} and \ref{fig:col_plots2}. These are the Tornado head (blue circle), north-western head (green dashed ellipse, south eastern head (pink dashed ellipse), Tornado tail (gold dashed ellipse), and the Tornado eye (white dash-dotted circle).
	}
	\label{fig:head_data}
\end{figure}

\begin{figure}
	\includegraphics[width=0.95\linewidth]{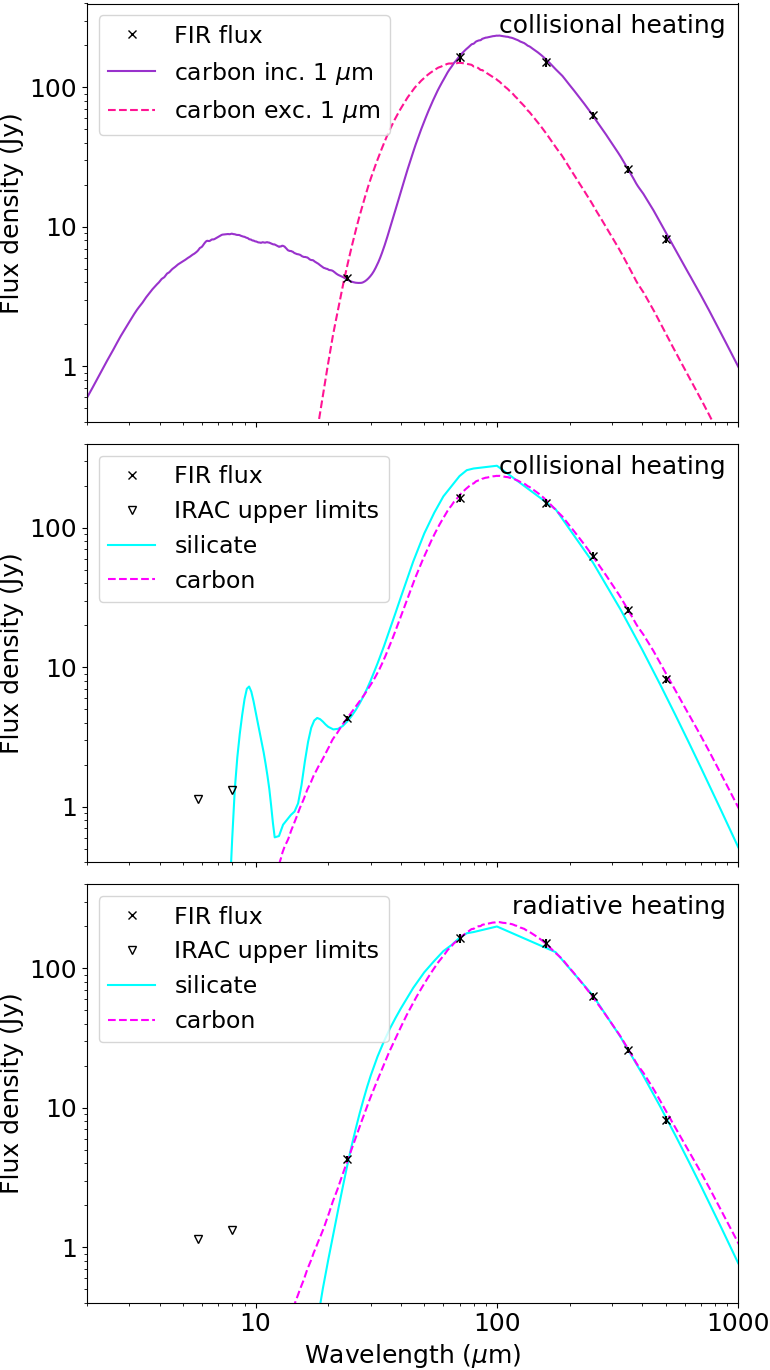}
	\caption{
	Best-fit dust SEDs for the Tornado head assuming that dust is collisionally heated by hot gas in the top two panels, and radiatively heated in the bottom panel.
	We use DINAMO \citep{Priestley2019} to fit to the flux within the head aperture in Fig.\ref{fig:head_data}, assuming the gas properties estimated by \citep{Sawada2011}.
	Although we can fit the SED well to the measured FIR fluxes with a collisional heating model and carbon grains, this requires a highly unusual grain size distribution. It is more likely that the majority of the dust within the Tornado head is radiatively heated, with a small proportion of collisionally heated dust.
	}
	\label{fig:tornado_sed_grains}
\end{figure}

In previous investigations both \citet{Sawada2011} and \citet{Gaensler2003} detected thermal X-ray emission from the head of the Tornado. This led \citet{Gaensler2003} to suggest that the head is a mixed-morphology SNR, centrally filled with thermal X-ray emission from shocked gas. In Figs.~\ref{fig:tornado_temps_ppmap} and \ref{fig:tornado_temps_3col_ppmap} we find that the warmest dust ($\sim$60~K) is at the location of the \xmm\ X-ray peak, thus we investigate whether the dust in the head is likely to be collisionally heated by hot, shocked gas.

We calculate grain temperatures and corresponding emissivities for grain sizes between 0.001\,--\,1~\micron\ using DINAMO \citep{Priestley2019}, a dust heating code which takes into account temperature fluctuations of small grains. We assume that the dust is heated by gas with the properties measured by \citet{Sawada2011} ($kT = 0.73$~keV and $n_e = 0.49$~cm$^{-3}$), and use optical properties for either BE amorphous carbon \citep{Zubko1996} or MgSiO$_3$ grains \citep{Dorschner1995}. The corresponding opacities at 300~\micron\ are 0.79~m$^2$g$^{-1}$ and 0.32~m$^2$g$^{-1}$ ($\beta =$ 1.5 and 1.7) respectively. The minimum equilibrium grain temperature, for micron-sized grains of either composition, is $\sim$30~K, so no set of grain properties result in an emissivity resembling a 20\,--\,30\,K blackbody, as indicated by PPMAP.

Following the method used in \citep{Priestley2020}, we fit the IR SED to background-subtracted fluxes within the blue head aperture in Fig.~\ref{fig:head_data} using a combination of single-grain SEDs for radii of 0.001, 0.01, 0.1 and 1~\micron\, with the number of grains (or equivalently the dust mass) of each size as the free parameters. We are unable to fit the FIR fluxes if we exclude 1~\micron\ grains. For carbon grains, shown in the top panel of Fig.~\ref{fig:tornado_sed_grains}, even 0.1~\micron\ grains have a 24/70~\micron\ ratio which is larger than the observed value, while at longer wavelengths the discrepancy becomes even more extreme. Silicate grains have the same issue, to a slightly greater extent. With 1~\micron\ radius grains included, we are able to reproduce the SED well at all wavelengths. We include IRAC fluxes (which may have significant non-SN dust contamination) as upper limits, in order to better constrain the number of transiently heated small grains, and we find best-fit dust masses of 8.1~M$_\odot$ for carbon grains and 17.3~M$_\odot$ for silicates. The best-fit SEDs are shown in Fig.~\ref{fig:tornado_sed_grains}.

In order to fit the FIR fluxes, both carbon and silicate grains require the vast majority ($\sim$99\,per\,cent) of the dust mass to be in micron-sized grains, while also requiring 0.05\,--\,0.06 M$_\odot$ of small grains with $a\leq$0.01~\micron\ to reproduce the 24~\micron\ emission. The mass of intermediate-sized grains with radius 0.1~\micron\ is strongly constrained to be below 10$^{-4}$~M$_\odot$, where they have a negligible contribution to the total SED. This distribution of grain sizes is highly unusual, both for the high mass fraction of micron-sized dust - the \citet{Mathis1977} power law does not extend to 1~\micron\, and even if extended results in only $\sim$30\,per\,cent of the mass in the largest grains - and the `bimodal' distribution of small and large grains.
Additionally, assuming a gas to dust ratio of 100, a dust mass of $\sim$10\,M$_\odot$ implies a gas mass of $\sim$1000\,M$_\odot$, much larger than that indicated by the X-ray emission \citep[M$_{gas}= $23\,M$_\odot$,][]{Sawada2011}.
We consider it more probable that the assumption of all grains being heated by the X-ray emitting gas is wrong. The synchrotron radiation generated by the shocked gas will heat nearby grains, both in the unshocked ISM and in any local over-densities which survive the blast wave, potentially resulting in a population of grains at lower temperatures.

While fully investigating the potential range of spectral shapes and intensities is beyond the scope of this paper, we can approximate it by scaling the \citet{Mathis1983} radiation field by a constant factor G. Assuming that the radiatively heated dust follows an MRN size distribution, we are able to fit the SED without the addition of micron-sized grains for G\,=\,5 for carbon and 10 for silicates. The best-fit SEDs, shown in Fig.~\ref{fig:tornado_sed_grains}, require 9.1~M$_\odot$ and 0.33~M$_\odot$ of radiatively and collisionally heated dust respectively for carbon grains. The size distribution of the collisionally heated dust is also reasonable, with the majority of the mass at 0.1~\micron\ and a negligible fraction of 0.001~\micron\ grains, as would be expected from an initial size distribution affected by sputtering \citep{Dwek1996}. For silicates, the radiatively and collisionally heated dust masses are 35.7 and 0.76~M$_\odot$ respectively. We note that these dust masses are not authoratative - differences in the assumed grain properties, size distribution and radiation field could cause significant variation in the best-fit masses. However, it is clear that a moderately-enhanced radiation field in the vicinity of the Tornado, combined with a small mass of dust in the shocked plasma, can explain the observed IR SED without any additional assumptions.
Our G\,=\,6 carbon model has a total cold dust luminosity of 2.6$\times$10$^{37}$erg\,s$^{-1}$ which can be explained by radiative heating via synchrotron radiation from the shock wave, given $\alpha=-0.63$ \citep{Law2008}.
We consider this explanation much more reasonable than invoking an arbitrary, and somewhat unphysical, size distribution for the dust in the hot plasma.
Investigations of the IR-X-ray flux ratio may give a more detailed description of the processes within the Tornado head, as shown for other SNRs by \citet{Koo2016}, although possible absorption by dense gas and molecular material in the vicinity makes this complicated.

In Section~\ref{subsec:ppmapResults} we estimated that the head of the Tornado contains a large dust mass of \ppmapDustMass\,M$_\odot$. This is unexpected for the mass within a SNR.
However, if the Tornado head is a SNR, it will have swept up a large mass of dust from the ISM through expansion. Assuming a simple relation where the swept up mass is equal to $\frac{4}{3} \pi R^3 \rho$, with a standard ISM density for cool, dense regions of $\rho=10^{-21}$\,kg\,m$^{-3}$, this gives a mass of $\sim5.26$\,M$_\odot$. As the ISM in this region is expected to be relatively dense, the swept up mass will likely be larger than this; assuming a gas density of $10^4$\,cm$^{-3}$ \citep{Lazendic2004} and dust-to-gas ratio of 100, the total swept up dust mass could be as large as $\sim$250\,M$_{\odot}$. 
Therefore, the dust mass of the Tornado head can be explained by material which has been swept up by an expanding SNR.

\section{The Nature of the Tornado} \label{sec:nature}
The nature of the Tornado is unclear as it has many confusing characteristics, with suggested candidates including an X-ray binary, a SNR, and \HII\ region.
In Section\,\ref{sec:methods} we revealed that the Tornado contains large masses of dust, similar to the sandy whirlwind `Dust Devils' on Earth. In this section we explore whether the FIR emission from our own Dust Devil can give us any insight into its nature. We further examine the IR, radio, and X-ray emission to determine if it can shine any light on the different origin scenarios.

\subsection{Properties of the Tornado} \label{sec:properties}
\begin{figure*}
	\includegraphics[width=0.75\linewidth, trim = 0cm 0cm 0cm 0.6cm, clip]{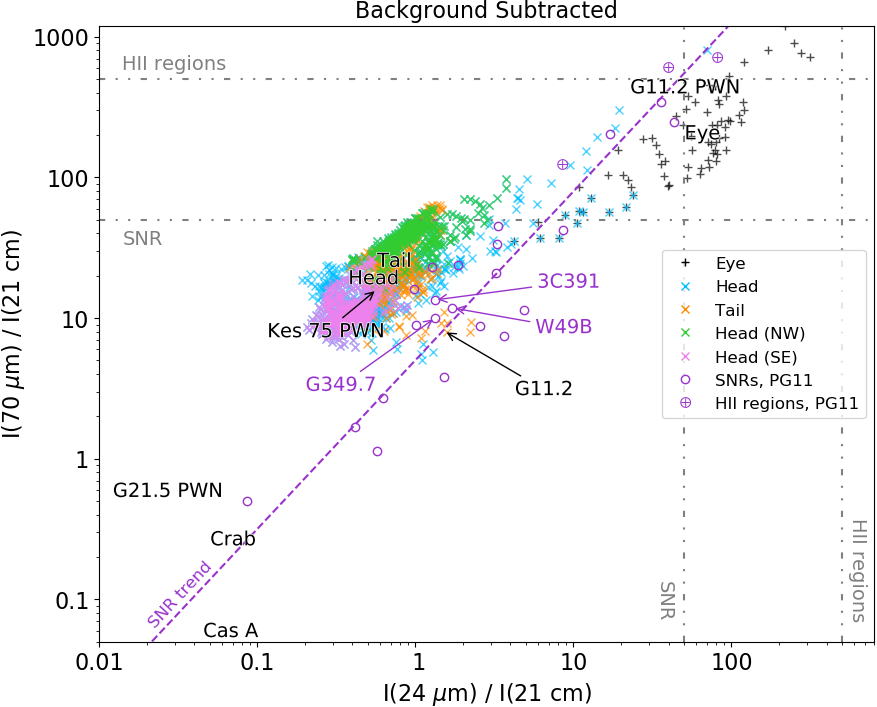}
	\caption{Flux ratio of individual pixels and integrated flux within the Tornado head, eye, and tail (within the circled regions in Fig.~\ref{fig:head_data}), in comparison with other SNRs and \HII\ regions. Pixels with very low signal have been removed, where the signal divided by the subtracted background is <0.1.
	The fluxes for the NW and SE head, and the tail are measured from the regions indicated in Fig.~\ref{fig:head_data}.
	The text labels are centred on the integrated flux for the Tornado head and eye, and previously studied SNRs, estimated by \citet[Cas A and Crab;][]{DeLooze2017, DeLooze2019} and \citet[G11.2, G21.5, G29.7 and G351.2;][]{Chawner2019, Chawner2020}.
	The grey dashed-dotted lines indicate ratios of 50 and 500, used in previous studies as diagnostics of SNRs and \HII\ regions.
	The majority of the Tornado head and tail pixels fall within the SNR region, and are clearly different to the pixels within the eye, which sits very close to the \HII\ region area of the colour space.
	All regions of the Tornado are found towards the upper right of the SNR regions, suggestive of an older remnant. There is a noticeable variation in the flux ratio of the NW and SE regions of the head.
	}
\label{fig:col_plots1}
\end{figure*}	

\begin{figure*}
	\includegraphics[width=0.75\linewidth]{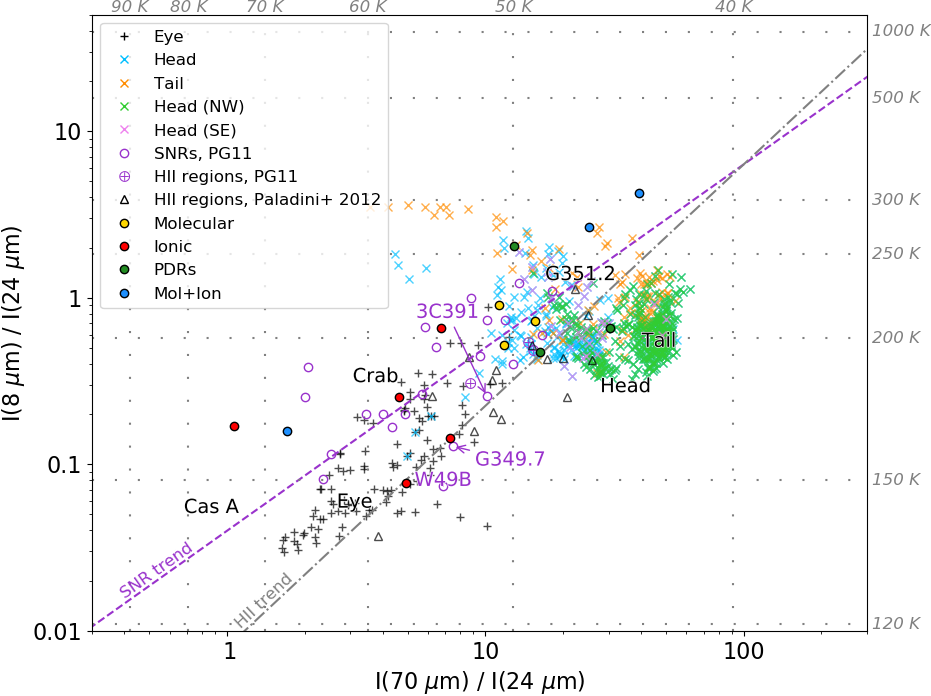}
	\caption{
	Flux ratio of individual pixels within the head, eye, and tail of the Tornado (within the circled regions in Fig.~\ref{fig:head_data}), in comparison with other SNRs and \HII\ region. Pixels with very low signal have been removed, where the signal divided by the subtracted background is <0.1. 
	The fluxes for the NW and SE head, and the tail are measured from the regions indicated in Fig.~\ref{fig:head_data}.
	The text labels are centred on the integrated flux for the Tornado head and eye, and previously studied SNRs, estimated by \citet[Cas A and Crab;][]{DeLooze2017, DeLooze2019} and \citet[G11.2, G21.5, G29.7 and G351.2;][]{Chawner2019, Chawner2020}.
	We also include ratios for SNRs with known molecular interactions, atomic  fine-structure emission, and PDRs from \citet{Goncalves2011}.
	The purple dashed and the grey dash-dotted lines indicate SNR and \HII\ region trends respectively, found by \citet{Goncalves2011}. SNRs populate a wider area in this colour space and several \citet{Goncalves2011} SNR measurements lie along the \HII\ region trend, including those highlighted in pink text.
	The grey dotted lines show the flux ratios expected from a thermal source with $\beta=2$ and the temperature indicated.
	The Tornado is found towards the upper right of this colour space, suggestive of an older remnant. It is also found in a region populated mainly by SNRs with molecular interactions.
}
\label{fig:col_plots2}
\end{figure*}

First, we study the emission colours to understand the properties of the regions from which we detect dust and how they vary across its features.
Within the head, we split our analysis into two main regions of interest, as indicated by the green and magenta ellipses in Fig.~\ref{fig:head_data} respectively: the north-west (NW), where we identified warm dust with \ppmap\ and where the head is thought to be interacting with a molecular cloud \citep{Frail1996,Lazendic2004,Hewitt2008}, and the south-east (SE), where there is a radio peak.
Our PPMAP analysis in Section~\ref{sec:methods} gives estimates for the dust mass within each of these regions as $\sim3.3 M_{\odot}$ and $\sim2.1 M_{\odot}$ for the NW and SE respectively.

IR\,--\,radio flux ratios have been used in previous studies to identify SNRs, distinguishing from \HII\ regions \citep[e.g. ][]{Whiteoak1996}.
The thermally dominated emission from \HII\ regions, with some free-free emission in the radio, gives an IR-radio ratio of $\geqslant$\,500; in contrast, SNRs are dominated by synchrotron at radio frequencies and have a considerably smaller IR flux, giving an IR\,-\,radio ratio of $\leqslant$\,50 \citep{Haslam1987, Furst1987, Broadbent1989}.

In order to examine the dust emission properties of the various FIR regions of the Tornado, we follow the analysis of \citet{Goncalves2011} and compare IR and radio colours, including $I_{\rm 70\mu m}/I_{\rm 21cm}$, $I_{\rm 24\mu m}/I_{\rm 21cm}$, $I_{\rm 8\mu m}/I_{\rm 24\mu m}$ and $I_{\rm 70\mu m}/I_{\rm 24\mu m}$, for pixels within the Tornado (Figs.~\ref{fig:col_plots1} and \ref{fig:col_plots2}), where pixels are convolved to the lowest resolution data. For comparison we include the integrated flux of the head, the dusty region in the tail (see Fig.~\ref{fig:head_data}), the eye, and previously studied SNRs (in Figs.~\ref{fig:col_plots1} and \ref{fig:col_plots2} the SNR and region names are centred on the respective flux ratios, unless indicated by an arrow).

In Fig.~\ref{fig:col_plots1} we find that the IR colours for the majority of the Tornado head pixels fall within the colour space for a SNR, and are well distinguished from the pixels within the `eye' of the Tornado, which is a confirmed \HII\ region with an embedded protostellar source \citep{Burton2004}. This suggests that the Tornado head is part of a SNR, rather than a \HII\ region. Several Galactic SNRs from \citet{Goncalves2011} are observed to have high IR-radio flux ratios, two of which would be classified as \HII\ regions by this test ($I_{\rm IR}/I_{\rm radio}>500$: G21.5$-$0.1 and G23.6$+$0.3, $I_{\rm IR}/I_{\rm radio}>50$: G10.5$+$0.0, G14.3$+$0.1, G18.6$-$0.2, and G20.4$+$0.1). Of these sources, \citet{Anderson2017} suggested that three were misidentified \HII\ regions (G20.4, G21.5, and G23.6), which we have labelled in Figs.~\ref{fig:col_plots1} and \ref{fig:col_plots2}.

As shown in  Fig.~\ref{fig:col_plots2}, \citet{Goncalves2011} found different trends for \HII\ regions and SNRs when comparing their IR colours. In this colour space, we find that the Tornado falls more in line with the \HII\ region trend. However, SNRs and \HII\ regions inhabit much of the same colour space in Figs.~\ref{fig:col_plots1} and \ref{fig:col_plots2} and there are other well known SNRs, including W49B, 3C391, and G349.7$-$0.2, which also lie along the \HII\ region trend. The variation seen in these individual SNRs from the main SNR trend could instead be due to a difference in dust properties such as temperature or emissivity, possibly caused by interactions with molecular clouds.

It is possible to use the IR and IR\,--\,radio colours, as in Figs.\,\ref{fig:col_plots1} and \ref{fig:col_plots2}, to determine some of the SNR properties. Older SNRs tend to have higher IR\,--\,radio colours \citep[e.g.][]{Arendt1989}, placing them towards the upper-right of the SNR colour space in Fig,\,\ref{fig:col_plots1}. Additionally, \citet{Goncalves2011} found some correlation between the IR colours in Fig.\,\ref{fig:col_plots2} and the SNR age, suggesting that older remnants have higher 70\,--\,24\,\micron\ and 8\,--\,24\,\micron\ flux ratios.
Thus, both the FIR\,--\,radio and the IR colours suggest that, if it is a SNR, the Tornado is an older remnant which has likely swept up a large mass of dust from the ISM.
\citet{Goncalves2011} also suggested that the IR colours could give some insight into the SNR emission process. They found tentative evidence that the upper right region of the colour space in Fig.\,\ref{fig:col_plots2} tends to be populated by objects with molecular shock and photodissociation regions (PDRs), although they admit that this is not a secure correlation given the small sample and that the 8, 24, and 70~\micron\ bands may contain both dust emission and lines. We find that the IR flux ratios of the NW region of the Tornado Head suggests molecular emission, whereas the SE region is largely undetected at 8\,\micron.
Given that the head is thought to be interacting with a molecular cloud in the NW, this supports the relation between the 70~\micron\,--\,24~\micron\ and 8~\micron\,--\,24~\micron\ flux ratios and emission type.

In all of the colour plots we find that the NW and SE regions (Fig.\ref{fig:head_data}) of the head are distinct and must have different emission processes.
Fig.~\ref{fig:col_plots1} shows a higher FIR\,--\,radio flux ratio in the NW region, suggesting an increased amount of thermal emission in the same area in which we see warm dust in Fig.~\ref{fig:tornado_temps_ppmap}: this dust may be heated through an interaction on this side.

\subsection{What the Devil is it?} \label{subsec:nature}
\citet{Gaensler2003} found that the X-ray emission from the head can be well explained by thermal models, rather than synchrotron emission, with a gas temperature of $kT \sim 0.6$\,keV, arising from the interior of a limb-brightened radio SNR.
Indeed, in Fig.~\ref{fig:tornado_temps_3col_ppmap} we find that the warmest dust is coincident with X-ray emission in the central region where hot gas may be heating the dust, as expected for a mixed-morphology SNR \citep{Rho1998, Yusef-Zadeh2003}.
\citet{Sawada2011} estimated an X-ray temperature of 0.73 keV for the head. Using an X-ray temperature of 0.73 keV (T\,=\,8.6  where T is in a unit of 10$^6$\,K) and assuming that the Tornado nebula is an SNR, we estimate a shock velocity (V$_s$) and age ($t$) of the SNR using the radius of only the head and both the head and tail (1.3$^\prime$ and 5.4$^\prime$). The shock velocity is 884 km/s based on $V_s = (T/11.)^{0.5}\times1000$\,kms$^{-1}$ \citep{Winkler1974}. The age of the SNR ($t = 2/5 Rs/Vs$) is therefore between 2000 and 8000\,yr. 

The bizarre shape of the tail is more difficult to explain with a SNR scenario.
\citet{Gaensler2003} suggested the tail could be explained by a progenitor star moving across the space whilst losing mass, which then exploded as a SN at the edge of circumstellar material \citep[CSM; ][]{Brighenti1994}. 
A similar scenario has been suggested for the SNR VRO\,42.05.01 \citep[G166.0$+$4.3,][]{Derlopa2020} which is much larger than the Tornado but morphologically resembles the Tornado head and surrounding filaments.
When a progenitor star  moves in relatively higher density interstellar medium (ISM), the stellar motion could cause a bow shock at the site of interaction between CSM and ISM.
Bow shocks have been detected in the red-supergiants $\alpha$ Ori and $\mu$ Cep \citep{NoriegaCrespo1997,Martin2007,Ueta2008,Cox2012}.  In the former, the bow shock has a wide opening angle, whereas the latter has a narrow-angle cylinder-type bow shock.  The cylinder shape of the Tornado's tail could therefore be explained by CSM-ISM interaction.
However, the CSM from red-supergiants does not emit synchrotron emission, so that the radio emission observed in the Tornado's tail would require additional energy by the SN-CSM interaction.
This requires the SN explosion itself to be highly elongated with very fast blast winds towards the east by more than by a factor of 10 to the west, which is unlikely and not supported by the hydrodynamic model \citep{Brighenti1994}.
Instead of synchrotron, the radio tail emission could be free-free; however, in that case, there should be some major heating and an obvious ionising source in the tail, which we do not see in the {\it Spitzer} 24~$\mu$m image (Fig.~\ref{fig:G357.7-0.1Image}).
Instead of a red-supergiant, the progenitor star could be a Wolf Rayet (WR) star, which has ionised gas in the CSM, and hence can emit free-free emission at radio wavelengths.
However, the lifetime of a WR star is too short to form such a large scale structure while the star is moving in the local space. The typical lifetime of a WR star is 10\,--\,36~kyrs \citep{Meynet2003,Meynet2005}.
At a distance of 12\,kpc, the furthest filament (centred at approximately $\alpha = 17^\text{h}40^\text{m}43.8^\text{s}, \delta = -30^\circ55^\prime44.9''$) is $\sim25$\,pc from the centre of the Tornado head.
This requires a progenitor to move through the ISM at speeds of approximately 1,000\,km\,s$^{-1}$. Though not impossible, such a high speed motion is unlikely. It is therefore difficult to explain the Tornado's tail with past mass loss from a SN (SN-CSM interaction).

Although the X-ray and radio emission from the head can be explained by thermal and synchrotron radiation from a SNR, the presence of an X-ray binary within the SNR would explain the length and the morphology of the tail in radio emission \citep{Helfand1985a, Stewart1994}.
\citet{Stewart1994} detected a spiral magnetic field around both the head and tail which they proposed could be explained by outflows from the central source dragging existing fields along the precession cone. In this instance, thermal X-ray emission at the location of the head is expected to arise from interactions between the jets and surrounding nebula, similar to that seen in the X-ray binary SS433 surrounded by the SNR W50 \citep{Brinkmann1996, Safi-Harb1997}.
The radio power law index of the central part of W50 is found to be typical for SNR \citep[$\alpha \sim$0.58,][]{Dubner1998}, while a hydrodynamic model shows that episodic jets from an X-ray binary containing a black hole compresses the SNR shell, forming a cylinder/helical shaped outflow in one direction \citep{Goodall2011}.

If the Tornado is formed by a binary system, the location of its source is controversial. In the case of the W50\,--\,SS433 system, the high mass X-ray binary is located in the SNR, following which would place the Tornado binary within the head.
However, \citet{Sawada2011} suggested that there is a \suz\ 1.5\,--\,3.0\,keV band detection of a `twin' source, opposite to where X-ray emission is already detected in the Tornado head. They propose that this originates from the interaction between the second jet of an X-ray binary system and a molecular cloud, placing any potential binary system source at the middle of the structure seen in Fig.~\ref{fig:G357.7-0.1Image}, rather than in the head. In this case, one might expect visible emission in the IR/FIR wavelengths at the location of the `twin' due to shocked gas/heated dust arising from jet interaction with the ISM. In the 24~\micron\ and the \hersc\ bands there is emission towards the south-west of this region which correlates with radio structures in the tail.
However, we do not see any clear evidence for an IR counterpart of the `twin': in all \spitz\ and \hersc\ maps the flux at the location of the \suz\ peak is at a similar level to, or lower than, that of the surrounding area (see Fig.\,\ref{fig:TwinMultiband}).
There is some X-ray emission in the \xmm\ and \chan\ data at the location of the `twin', although the emission does not seem correlated. However, the X-ray emission may be affected by foreground absorption, making association difficult to determine, and the region may peak in the 1.5\,--\,3.0\,keV \suz\ band with much lower emission of softer X-ray, making comparison between multiple bands complicated. 

As there does seem to be X-ray and radio emission at the location of the `twin' it is plausible that there is an object in this region, which may be associated with the Tornado as suggested by \citet{Sawada2011}. However, if there is emission from such an object in any of the \spitz\ or \hersc\ bands, it is very faint and is not detected above the level of the ISM in this region (Fig.~\ref{fig:TwinMultiband}). This is unlike the head, from which there is a clear detection in the 5.8\,--\,500~\micron\ bands, as well as a very bright radio structure (Fig.~\ref{fig:HeadMultiband}). It seems strange that their IR profiles are so different if the two regions have been formed by a similar process, although we cannot exclude this as a possibility.
If the X-ray `twin' head is unrelated to the Tornado, it is plausible that the location of an X-ray binary, if any, could be within the head of the Tornado as discussed above.

Although the IR-radio emission supports a SNR origin for the Tornado head, we see no clear indication that the X-ray emission from the head results from an interaction between X-ray binary jets and the surrounding nebula. However, the helical shape of the tail, and the presence of its magnetic field and synchrotron radiation, can be explained by a jet ploughing into a SNR shell, as observed in W50. Although there is no detection of a central powering source, there are cases in which the central X-ray binary may be too faint to detect at a distance of 12\,kpc. \citet{Gaensler2003} suggest that this would be the case for a high-mass X-ray binary such as LS\,5039 \citep{Paredes2000}, from which the luminosity may vary with orbital phase and its minimum is slightly higher than the upper limit for detection of a Tornado central source. It could also be the case that the Tornado is powered by a low mass X-ray binary in a quiescent state, having produced the observed features in a past period of prolonged activity \citep{Sawada2011}, as seen in 4U\,1755--338 \citep{Angelini2003}.

\section{Conclusion} \label{sec:Summary}
We detect FIR emission from dust in the unusual SNR candidate the Tornado (G357.7$-$0.1), akin to the terrestrial sandy whirlwinds known as `Dust Devils'. We investigate the distribution of dust in the Tornado using Point Process Mapping, \ppmap.
Similar to that found in the radio emission, we find a complex morphology of dust structures at multiple temperatures within both the head and the tail of the Tornado, ranging from 20\,--\,60\,K.  In the head of the Tornado, we find warm dust in the region at which the object is thought to be interacting with a molecular cloud. We also find a filament along the SE edge coinciding with radio emission, and a cool dusty shell encapsulating hot dust near to the location of an X-ray peak.
We derive a total dust mass for the head of the Tornado of \ppmapDustMass\,$\rm M_{\odot}$, and we find that the majority of the dust is most likely heated radiatively, with a small proportion of collisionally heated dust.
When considering that the Tornado may be a SNR, we find that it is aged between 2000 and 8000\,yrs and it is plausible that the estimated dust mass originates from material swept up from the ISM.

The origin of the Tornado is still unclear.
We do not find clear evidence of a FIR counterpart to the Tornado `twin' detected by \citet{Sawada2011}, which was suggested to be the other end of an X-ray binary system.  The FIR-radio colours in the Tornado head are consistent with a SNR origin for this structure, yet the tail is not easily explained via just the SN or a SN-CSM interaction.  The tail can be explained via jets from an X-ray binary source within the nebula, similar to the W50 SNR.
One useful way to distinguish between the several hypotheses put forward by various authors would be to measure the velocity of the gas motion in the tail, if it emits in near-infrared Br $\alpha$ or [Fe II] for example.

\section*{Acknowledgements}
We thank the referee John Raymond for their helpful and constructive referee report.

We thank Bryan Gaensler for discussion on the nature of the Tornado and sharing X-ray \xmm\ data.

HC, HLG, and PC acknowledge support from the European Research Council (ERC) in the form of Consolidator Grant {\sc CosmicDust} (ERC-2014-CoG-647939).
MM acknowledges support from an STFC Ernest Rutherford fellowship (ST/L003597/1).
FP and AW acknowledges support from the STFC (ST/S00033X/1).
MJB acknowledges support from the ERC in the form of Advanced Grant SNDUST (ERC-2015-AdG-694520).
IDL gratefully acknowledges the support of the Research Foundation Flanders (FWO).
ADPH gratefully acknowledges the support of a postgraduate scholarship from the UK Science and Technology Facilities Council.
\hersc\ is an ESA space observatory with science instruments provided by European-led Principal Investigator consortia and with important participation from NASA.
JR acknowledges support from NASA ADAP grant (80NSSC20K0449).

This research has made use of data from the HiGAL survey (2012hers.prop.2454M, 2011hers.prop.1899M, 2010hers.prop.1172M, 2010hers.prop.358M) and Astropy\footnote{\url{http://www.astropy.org}}, a community-developed core Python package for Astronomy \citep{astropy2013,astropy2018}.
This research has made use of the NASA/ IPAC Infrared Science Archive, which is operated by the Jet Propulsion Laboratory, California Institute of Technology, under contract with the National Aeronautics and Space Administration.

\section*{Data Availability}
The \hersc\ and \spitz\ data underlying this article are available in Hi-GAL Catalogs and Image Server at https://tools.ssdc.asi.it/HiGAL.jsp and the Infrared Science Archive at https://sha.ipac.caltech.edu/applications/Spitzer/SHA/.
The VLA data are available in the National Radio Astronomy Observatory Science Data Archive at https://archive.nrao.edu/archive/advquery.jsp/
The \chan\ and \xmm\ data were provided by Bryan Gaensler by permission.
The \suz\ data are available in the Suzaku Archive at HEASARC at http://heasarc.gsfc.nasa.gov/docs/suzaku/aehp\_archive.html.

\bibliographystyle{mnras}
\bibliography{library}

\appendix

\section{Synthetic observations with \ppmap}
\label{app:ppmaptests}

\begin{figure}
	\includegraphics[width=0.95\linewidth, trim = 0.3cm 0.3cm 0.2cm 1.3cm, clip]{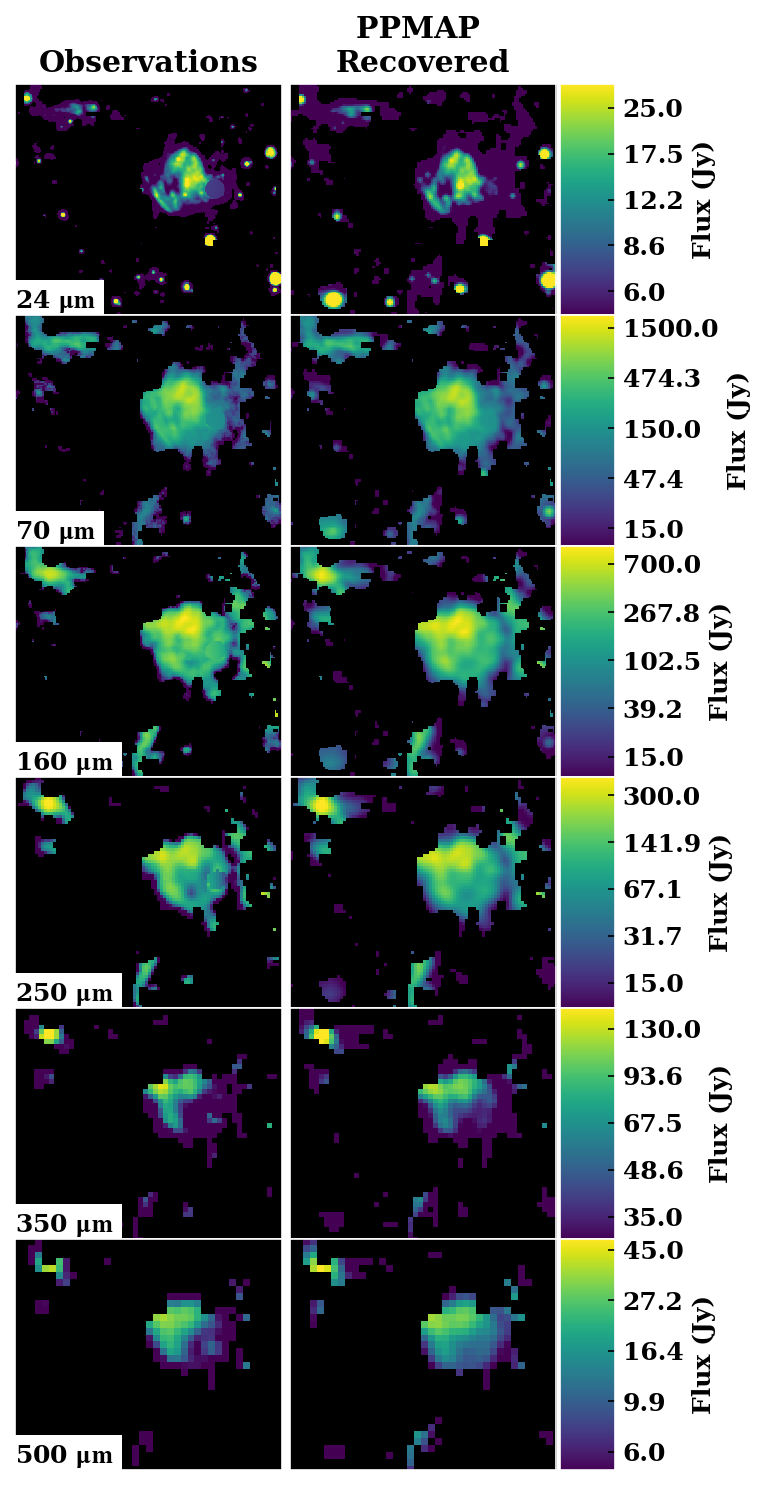}
	\caption{A grid comparing the original {\it Spitzer} and {\it Herschel} observations of the Tornado ({\it left}) with the synthetic observations ({\it right}) created by taking the results from \ppmap\ and post-processing them.
	}
	\label{fig:synthetic_ppmap}
\end{figure}

In order to try to quantitatively distinguish between the outputs based on different runs of \ppmap\ with different assumptions (and in particular using different estimates of background subtraction) we produced synthetic observations. These were created from the output dust column density maps at a range of temperatures and then reversing the physical steps \ppmap\ uses to produce maps of flux at each wavelength, ultimately regridding the pixels and smoothing back to the resolution of the original data. This also allows us to independently check no artefacts are introduced in \ppmap\ since these would be obvious in the synthetic images.  Fig.~\ref{fig:synthetic_ppmap} shows a comparison of the synthetic images from \ppmap\ versus the original data for the zero-mean-background-subtracted case. Here we see a close agreement with the dust structures and components seen in the head of the Tornado in the original data in all wavebands.

\section{The X-Ray twin of the head}
\label{app:twin}
\begin{figure*}
	\centering
	\includegraphics[width=0.8\linewidth]{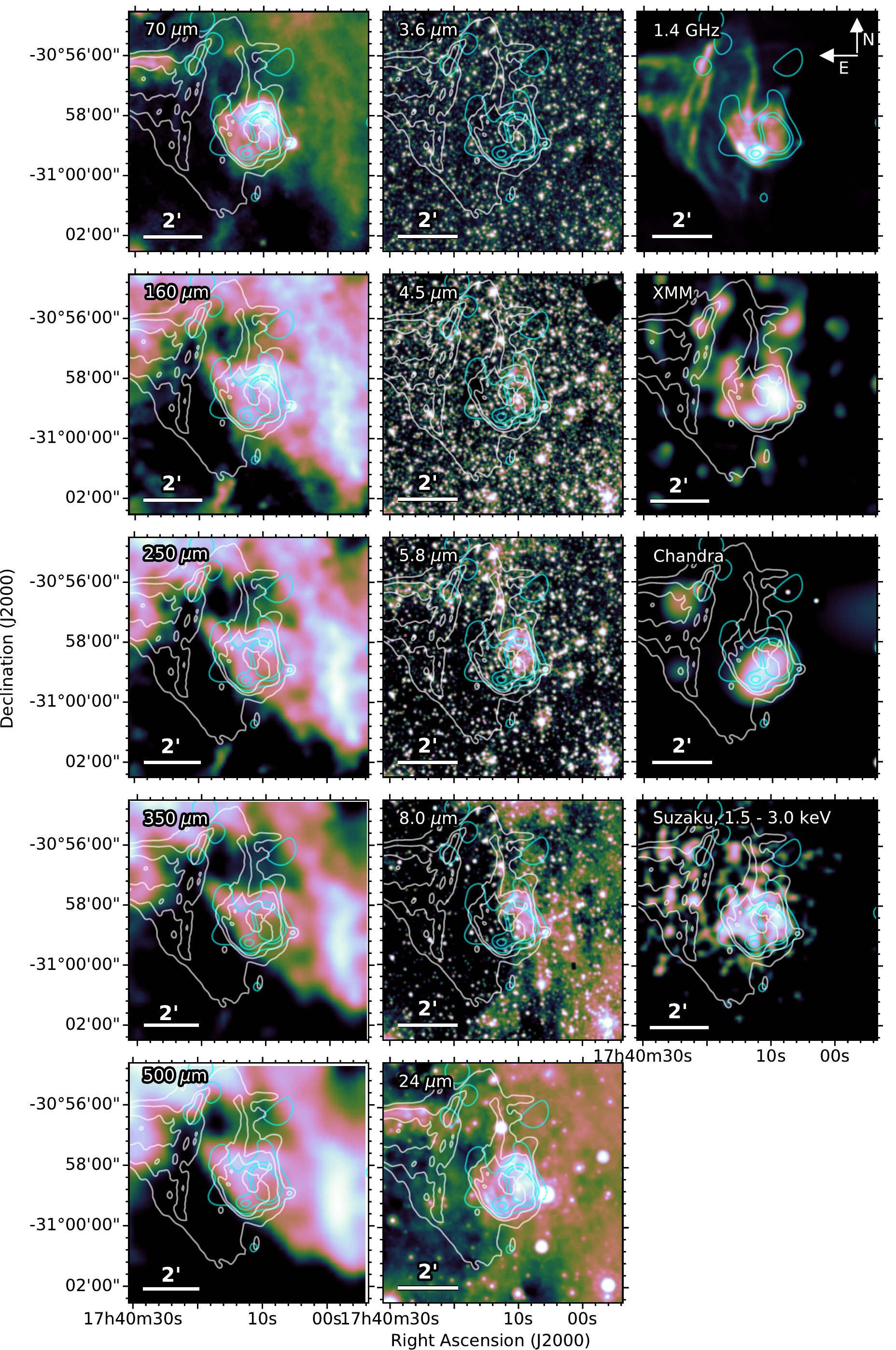}
	\caption{
	IR, radio, and X-ray view of the location of the Tornado head.
	{\it Left column:} \hersc\ images,
	{\it middle column:} \spitz\ images,
	{\it right top:} 1.4\,GHz VLA image,
	{\it right second row:} \xmm\ X-ray image, 
	{\it right third row:} \chan\ X-ray image, and
	{\it right bottom:} \suz\ 1.5\,--\,4.0\,keV X-ray image. We note that we have not applied a background subtraction or correction for vignetting as was done by \citet{Sawada2011}.
	The white and cyan contours show the VLA~1.4~GHz and \xmm\ emission respectively.
	There is a clear detection of emission from the head at the \spitz\ and \hersc\ wavebands, between 5.8 and 250\,\micron, at 3.6, 350, and 500\,\micron\ there is emission which seems associated although it is more confused. There is a clear detection in all of the radio and X-ray images.
	(We use the {\sc cubehelix} colour scheme, \citet{Green2011}.)
	}
	\label{fig:HeadMultiband}
\end{figure*}

\begin{figure*}
	\centering
	\includegraphics[width=0.8\linewidth]{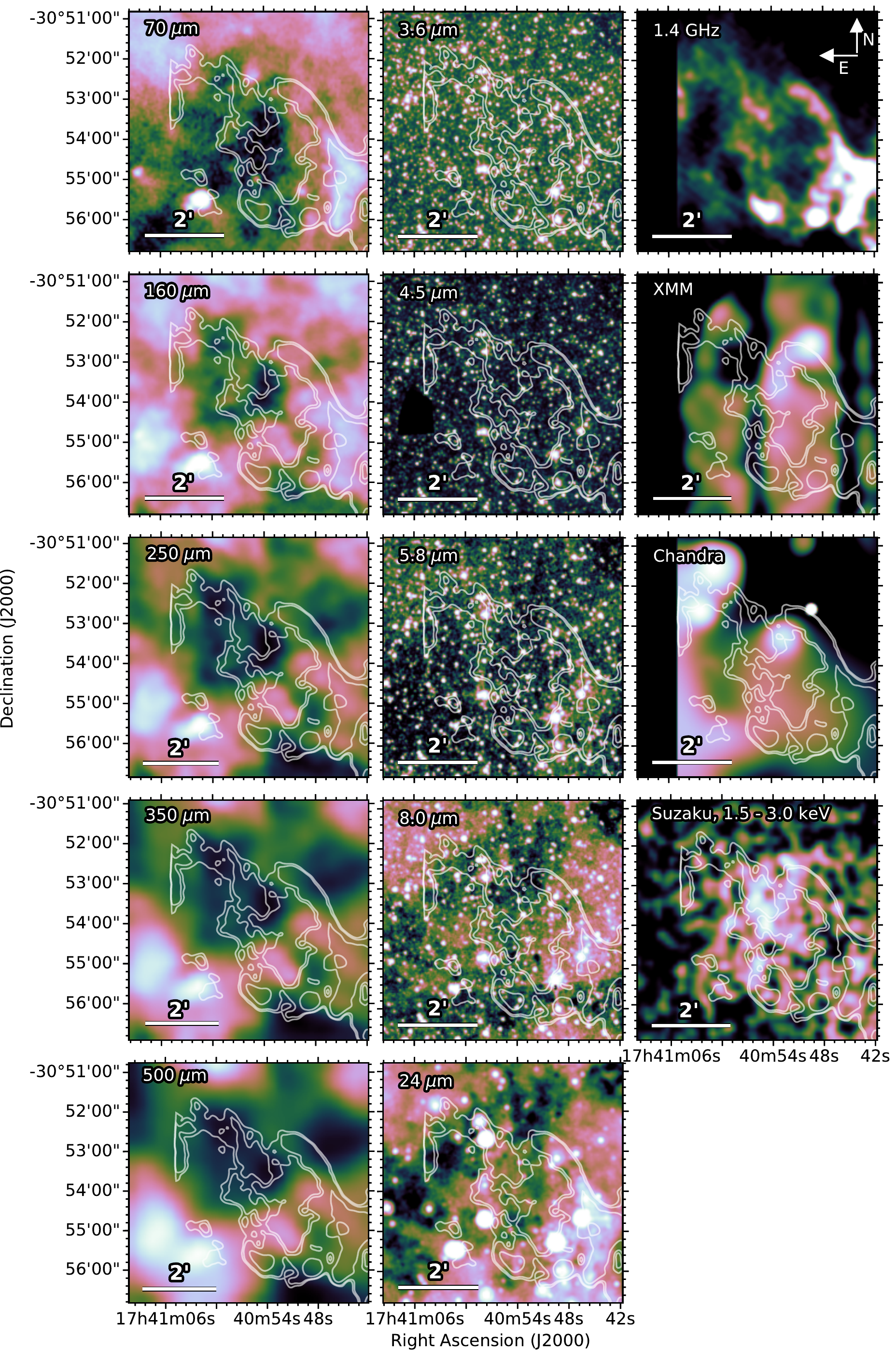}
	\caption{
	IR, radio, and X-ray view of the location of the X-ray twin, detected by \citet{Sawada2011}, the scale is increased compared with the image in Fig~\ref{fig:G357.7-0.1Image} to enhance any features in the region.
	{\it Left column:} \hersc\ images,
	{\it middle column:} \spitz\ images,
	{\it right top:} 1.4\,GHz VLA image,
	{\it right second row:} \xmm\ X-ray image, 
	{\it right third row:} \chan\ X-ray image, and
	{\it right bottom:} \suz\ 1.5\,--\,4.0\,keV X-ray image. We note that we have not applied a background subtraction or correction for vignetting as was done by \citet{Sawada2011}.
	The white contours show the VLA 1.4\,GHz emission.
	In all \hersc\ and IRAC bands the flux level at the location of the twin is similar to, or lower than, that of the surrounding ISM. In all other bands there is some emission, although the morphology is not consistent with the \suz\ features, and at 24\,\micron\ this is fainter than much of the surrounding ISM.
	(We use the {\sc cubehelix} colour scheme, \citet{Green2011}.)
	}
	\label{fig:TwinMultiband}
\end{figure*}

\bsp	
\label{lastpage}
\end{document}